\documentclass[preprint]{revtex4}

\usepackage{graphicx}
\usepackage{dcolumn}
\usepackage{bm}
\usepackage{color}
\usepackage{amsmath,amsthm}
\usepackage{amssymb,bm}
\usepackage{mathrsfs}
\usepackage{comment}
\usepackage{ulem}

\graphicspath{%
    {converted_graphics/}
    {/}
}
\begin{document}

\title{Controlling the nonadiabatic electron-transfer reaction rate through molecular-vibration polaritons in the ultrastrong coupling regime}
\author{Nguyen Thanh Phuc}
\email{nthanhphuc@ims.ac.jp}
\affiliation{Department of Theoretical and Computational Molecular Science, Institute for Molecular Science, Okazaki 444-8585, Japan}
\affiliation{Department of Structural Molecular Science, The Graduate University for Advanced Studies, Okazaki 444-8585, Japan}
\author{Pham Quang Trung}
\affiliation{Section of Brain Function Information, Supportive Center for Brain Research, National Institute for Physiological Sciences, Okazaki 444-8585, Japan}
\author{Akihito Ishizaki}
\affiliation{Department of Theoretical and Computational Molecular Science, Institute for Molecular Science, Okazaki 444-8585, Japan}
\affiliation{Department of Structural Molecular Science, The Graduate University for Advanced Studies, Okazaki 444-8585, Japan}

\begin{abstract}
Recent experiments showed that the chemical reaction rate is modified, either increased or decreased, by strongly coupling a nuclear vibration mode to the single mode of an optical cavity. Herein we investigate how the rate of an electron-transfer reaction depends on the molecule-cavity coupling in the ultrastrong coupling regime, where the coupling strength is comparable in magnitude with both the vibrational and the cavity frequencies. 
We found two main factors that determine the modification of the reaction rate: the relative shifts of the energy levels induced by the coupling and the mixing of the ground and excited states of molecular vibration in the ground state of the hybrid molecule-plus-cavity system through which the Franck-Condon factor between the initial and final states of the transition is altered. 
The former is the dominant factor if the molecule-cavity coupling strengths for the reactant and product states differ significantly from each other and gives rise to an increase in the reaction rate over a wide range of system's parameters. The latter dominates if the coupling strengths and energy levels of the reactant and product states are close to each other and it leads to a decrease in the reaction rate.
The effect of the mixing of molecular vibrational states on the reaction rate is, however, suppressed in a system containing a large number of molecules due to the collective nature of the resulting polariton, and thus should be observed in a system containing a small number of molecules. In contrast, the effect of the relative shifts of the energy levels should be essentially independent of the number of molecules coupled to the cavity.
\end{abstract}

\keywords{polariton, ultrastrong coupling, molecular vibration, reaction rate}

\maketitle

Controlling chemical reactions has always been an important goal in the field of chemistry. Over the last century, synthetic chemists have developed various kinds of catalysts to modify chemical reactions~\cite{Wender96, Carreira15}.
There are also physical methods to control chemical reactivity by using intense laser fields to excite the nuclear vibration to overcome the barrier of the reaction~\cite{Frei81, Sinha91, Zare98, Crim99}.
However, these approaches often require cryogenic temperature as the excitation energy can be redistributed to other vibrational degrees of freedom.
One way to overcome this challenge is to control the chemical reaction by strongly coupling the molecular vibration to the vacuum field of a cavity mode~\cite{Shalabney15, Long15, George16, Vergauwe16}. 
Strong coupling of both electronic and vibrational degrees of freedom of molecules to an optical cavity has already been realized in various experimental platforms involving both an ensemble of molecules~\cite{Ebbesen16, Lidzey98, Tischler05, Holmes07, Cohen08, Bellessa14, Long15, George15b, Shalabney15, George16, Vergauwe16} as well as a single molecule~\cite{Chikkaraddy16}. It gave rise to a variety of interesting phenomena and important applications including the control of chemical reactivity~\cite{Hutchison12, Simpkins15, Herrera16, Galego16, Thomas16, Thomas19, Hiura19, Lather19}, enhancement of transports~\cite{Hutchison13, Andrew00, Feist15, Schachenmayer15, Orgiu15, Zhou16}, nonlinear optical properties with applications to optoelectronic devices~\cite{Herrera14, Bennett16, Kowalewski16, Kowalewski16b}, polariton lasing and condensate~\cite{Cohen10, Cwik14, Lerario17, Plumhof14}, and precise measurement of molecular excitation energies~\cite{Phuc19}.

In particular, the interaction of the zero-point energy fluctuations of the cavity mode with the molecular vibration is expected to modify the chemical reactivity, which has recently been demonstrated experimentally~\cite{Thomas16, Thomas19, Hiura19, Lather19}. 
Surprisingly, in Refs.~\cite{Thomas16, Thomas19} the reaction rate is found to decrease whereas in Refs.~\cite{Hiura19, Lather19} the reaction rate increases if a molecular vibration is strongly coupled to the cavity mode. 
Theoretical investigations of the effect of the coupling between molecular vibrations and the optical cavity mode on the chemical reaction rate have also been done for the ab initio model of a simple molecule~\cite{Galego19} and the model of electron-transfer reaction~\cite{Angulo19}. 
However, the physical mechanism underlying the modification of the chemical reactivity induced by the molecule-cavity coupling is not fully understood.
In particular, the change of the ground state of the total system by the molecule-cavity coupling has been ignored when applying the rotating-wave approximation~\cite{Angulo19}. 
The mixing of ground and excited states of molecular vibration in the ground state of the hybrid system can, in principle, significantly affect the chemical reactivity in a way similar to the excitation of nuclear vibrations done by an intense laser. The only difference is that here the excitation is induced by the quantum fluctuation in the vacuum field of the cavity rather than by a strong laser field.

In this paper, we extend the investigation of the effect of vibrational polariton on the chemical reaction rate to the so-called ultrastrong coupling regime, where the molecule-cavity coupling strength is comparable in magnitude with both the vibrational and the cavity frequencies~\cite{Kockum19}. The ultrastrong coupling has already been realized in various kinds of systems including intersubband polaritons~\cite{Anappara09, Gunter09}, superconducting circuits~\cite{Niemczyk10, Yoshihara17}, Landau polaritons~\cite{Scalari12, Bayer17}, optomechanics~\cite{Benz16} as well as organic molecules~\cite{Schwartz11, Cohen13, Gambino14, Gubbin14, Mazzeo14, Todisco18, Barachati18, Eizner18}. In particular, the ultrastrong coupling between molecular vibrations and the cavity field has been realized in the experiment of Ref.~\cite{Hiura19}. 
In the ultrastrong coupling regime, the rotating-wave approximation is no longer valid and the optical diamagnetic term, which is proportional to $\hat{\mathbf{A}}^2$ where $\hat{\mathbf{A}}$ is the vector potential of the cavity field, cannot be neglected. As a result, the ground state of the total system can be strongly modified as it now involves virtual photons and molecule's vibrational excitations~\cite{Kockum19}.

In the following, we investigate how the ultrastrong coupling between molecular vibrations and an optical cavity can affect the electron-transfer reaction rate. 
There exist both regions of parameters for which the reaction rate increases or decreases by coupling the molecular vibration to the cavity field. 
The modification of the reaction rate by the molecule-cavity coupling is determined by two main factors: the relative shifts of the energy levels induced by the coupling, and the mixing of ground and excited states of molecular vibration in the ground state of the hybrid system through which the Franck-Condon factor between the initial and final states of the transition is altered.
The former is the dominant factor if the molecule-cavity coupling strengths for the reactant and product states differ significantly from each other. It increases the reaction rate over a wide range of system's parameters. 
Conversely, the latter dominates if the coupling strengths and energy levels of the reactant and product states are close to each other, and it counterintuitively leads to a decrease in the reaction rate. 
This is in contrast to the normal expectation that the reaction rate would increase due to the molecule's vibrational excitations. The result, however, can be understood as a consequence of the minus sign of the coefficient of the molecular vibration's excited state induced by coupling with the cavity.
We also investigate how the effect of vibrational polariton on the reaction rate changes for a variable number of molecules coupled to the cavity. 
The effect of the mixing of vibrational excitations on the reaction rate is suppressed in a system containing a large number of molecules due to the collective nature of the resulting polariton, while the effect of the relative shifts of the energy levels is essentially independent of the number of molecules.
It should be noted that in addition to the electron-transfer reaction~\cite{Marcus93, Piotrowiak98}, which plays a crucial role in various biological~\cite{Blumberger15, Beratan19} and chemical~\cite{Kaplan17, Mu19a, Mu19b, Mu19c} molecular systems including natural photosynthesis~\cite{Holzwarth06} and photoelectric functional materials~\cite{Bredas17}, the results obtained in this paper can also be applied to similar types of reactions including the excitation-energy transfer~\cite{Mirkovic17} and spin-singlet fission~\cite{Smith13} processes.

\section{Ultrastrong coupling between a molecular vibration and an optical cavity}
\label{sec: Ultrastrong coupling between a molecular vibration and an optical cavity}
\begin{figure}[tbp] 
  \centering
  \includegraphics[keepaspectratio]{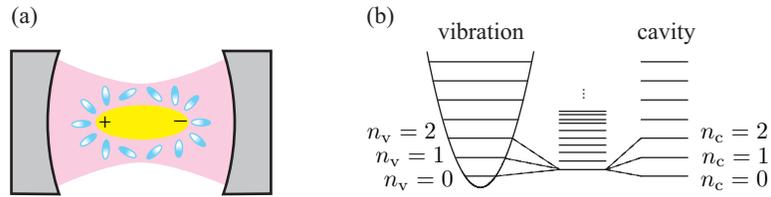}
  \caption{Control of electron-transfer reaction by ultrastrongly coupling a molecular vibration mode to a single mode of an optical cavity (magenta). Top: Schematic illustration of the configuration taken by the surrounding molecular environment (blue ellipsoids) in accordance with the charge distribution of the molecule (yellow ellipsoid). Bottom: Formation of energy eigenstates of the molecule-plus-cavity hybrid system by an ultrastrong coupling between the molecular vibration and the cavity mode. The ultrastrong coupling regime refers to the case that the coupling strength is comparable in magnitude with both the vibration and the cavity frequencies. The lowest-energy eigenstate of the hybrid system is a superposition of $|n_\mathrm{v}, n_\mathrm{c}\rangle$ states with $n_\mathrm{v}, n_\mathrm{c}=0, 1, 2, \cdots$ representing the numbers of vibrational quanta and photons, respectively.}
  \label{fig: system}
\end{figure}

We consider an electron-transfer chemical reaction in a system of $N$ identical molecules whose nuclear vibrations are coupled to an optical cavity.
As described by the Marcus-Levich-Jortner model~\cite{Marcus64, Levich66, Jortner76}, the reactant (R) and product (P) electronic states of each molecule are coupled to a single high-frequency vibrational mode, which is further coupled to a single mode of the optical cavity as illustrated in Fig.~\ref{fig: system}. The molecule's electronic states are also coupled to a continuum of low-frequency vibrational modes denoted by $\xi$, which stem from the inter-molecular vibrations of the surrounding molecular environment such as the solvent. The low-frequency vibrational modes are modeled by the annihilation operators $\hat{b}_\xi^{(i)}$ and the Hamiltonian $\hat{H}_\mathrm{e}^{(i)}=\sum_\xi \hbar \omega_\xi (\hat{b}_\xi^{(i)})^\dagger \hat{b}_\xi^{(i)}$ (for the $i$th molecule).

The total Hamiltonian of the molecular system is given by
\begin{align}
\hat{H}_\mathrm{m}= &\sum_{i=1}^N \Big[
\hat{H}_\mathrm{R}^{(i)} |\mathrm{R}_i\rangle \langle \mathrm{R}_i|
+\hat{H}_\mathrm{P}^{(i)} |\mathrm{P}_i\rangle \langle \mathrm{P}_i| \nonumber\\
&+\left(V_\mathrm{RP}  |\mathrm{P}_i\rangle \langle \mathrm{R}_i| + \mathrm{h.c.} \right)\Big] ,
\end{align}
where $V_\mathrm{RP}$ represents the coupling between the reactant and product states, and h.c. stands for Hermitian conjugate.
The Hamiltonians $\hat{H}_\mathrm{R}^{(i)}$ and $\hat{H}_\mathrm{P}^{(i)}$ are given by
\begin{align}
\hat{H}_\mathrm{R}^{(i)}=&
\hbar \omega_\mathrm{R} (\hat{a}_\mathrm{R}^{(i)})^\dagger  \hat{a}_\mathrm{R}^{(i)} + \Delta E_\mathrm{RP}
\end{align}
and 
\begin{align}
\hat{H}_\mathrm{P}^{(i)}=&
\hbar \omega_\mathrm{P} (\hat{a}_\mathrm{P}^{(i)})^\dagger  \hat{a}_\mathrm{P}^{(i)} 
+\sum_\xi g_\xi \left[ (\hat{b}_\xi^{(i)})^\dagger +\hat{b}_\xi^{(i)} \right] +\lambda.
\end{align}
Here, $\omega_\mathrm{R}$ and $\omega_\mathrm{P}$ are the frequencies of the harmonic potentials associated with the high-frequency vibrational mode near the equilibrium positions of the reactant and product states, respectively. The annihilation operators $\hat{a}_\mathrm{R}^{(i)}$ and $ \hat{a}_\mathrm{P}^{(i)}$ correspond to the nuclear vibrational motions in these harmonic potentials. The energy difference $\Delta E_\mathrm{RP}=E^\mathrm{R}_{n_\mathrm{R}=0}-E^\mathrm{P}_{n_\mathrm{P}=0}$ is between two vibrational ground states $|n_\mathrm{R}=0\rangle$ and $|n_\mathrm{P}=0\rangle$ associated with R and P. The coupling strength to the low-frequency $\xi$-mode is denoted by $g_\xi$ and $\lambda=\sum_\xi g_\xi^2/(\hbar\omega_\xi)$ is the reorganization energy with respect to the R-P transition. 

The operators $\hat{a}_\mathrm{R}^{(i)}$ and $ \hat{a}_\mathrm{P}^{(i)}$ of the two harmonic potentials are related to each other by $\hat{a}_\mathrm{P}^{(i)}=\hat{D}_i^\dagger \hat{S}_i^\dagger \hat{a}_\mathrm{R}^{(i)} \hat{S}_i \hat{D}_i$, where
\begin{align}
\hat{S}_i=&
\exp \left\{ \frac{1}{2} 
\ln \left( \sqrt{\frac{\omega_\mathrm{P}}{\omega_\mathrm{R}}} \right) 
\left[ (\hat{a}_\mathrm{R}^{(i)})^{\dagger 2} - (\hat{a}_\mathrm{R}^{(i)})^2 \right] \right\}
\end{align}
and
\begin{align}
\hat{D}_i=&
\exp \left\{ -\frac{d_\mathrm{RP}}{\sqrt{2}} 
\left[ (\hat{a}_\mathrm{R}^{(i)})^\dagger-\hat{a}_\mathrm{R}^{(i)} \right] \right\}
\end{align}
are the squeezing and displacement operators, respectively~\cite{May-book}. Here, $d_\mathrm{RP}$ is the dimensionless distance (normalized a factor proportional to $1/\sqrt{\omega_\mathrm{R}}$) between the two equilibrium positions of R and P along the high-frequency vibrational mode. For simplicity, however, in the following numerical calculations $\omega_\mathrm{R}=\omega_\mathrm{P}$ was set. 

The general coupling $\hat{\mathbf{p}}\cdot\hat{\mathbf{A}}$ between the molecular vibration and the single mode of the optical cavity is given in the Coulomb gauge by
\begin{align}
\hat{H}_\mathrm{int}=&
i\hbar \sum_{i=1}^N 
\Big\{ g_\mathrm{R} \left[ (\hat{a}_\mathrm{R}^{(i)})^\dagger - \hat{a}_\mathrm{R}^{(i)}\right] |\mathrm{R}_i\rangle \langle \mathrm{R}_i| \nonumber\\
&+g_\mathrm{P} \left[ (\hat{a}_\mathrm{P}^{(i)})^\dagger - \hat{a}_\mathrm{P}^{(i)}\right] |\mathrm{P}_i\rangle \langle \mathrm{P}_i| \Big\}\left( \hat{c}^\dagger +\hat{c} \right),
\label{eq: interaction Hamiltonian}
\end{align}
where $g_\mathrm{R}$ and $g_\mathrm{P}$ denote the coupling strengths for the reactant and product states, respectively.
Here, the momentum operator $\hat{\mathbf{p}}$ associated with the high-frequency vibrational mode is proportional to $i \left[ (\hat{a}_\mathrm{R}^{(i)})^\dagger - \hat{a}_\mathrm{R}^{(i)}\right]$ and $i \left[ (\hat{a}_\mathrm{P}^{(i)})^\dagger - \hat{a}_\mathrm{P}^{(i)}\right]$, while the vector potential operator $\hat{\mathbf{A}}$ is proportional to $\hat{c}^\dagger +\hat{c}$, where $\hat{c}$ is the cavity field operator. 

It is noteworthy that in the ultrastrong coupling regime under consideration, where the coupling strength $g_\mathrm{R,P}$ is comparable in magnitude with both the molecule's vibrational frequency $\omega_\mathrm{R,P}$ and the cavity frequency $\omega_\mathrm{c}$, the rotating-wave approximation is no longer valid and the counter-rotating terms $\hat{a}_{\mathrm{R},\mathrm{P}}^{(i)}\hat{c}$ and $(\hat{a}_{\mathrm{R},\mathrm{P}}^{(i)})^\dagger \hat{c}^\dagger$ must be considered as they are included in Eq.~\eqref{eq: interaction Hamiltonian}.
Moreover, in the ultrastrong coupling regime, the optical diamagnetic term proportional to $\hat{\mathbf{A}}^2$ also becomes comparable in magnitude with the light-matter interaction, and thus cannot be neglected~\cite{Nataf10, Liberato14, Kockum19}. Consequently, the total photonic Hamiltonian is given by
\begin{align}
\hat{H}_\mathrm{ph}=&
\hbar \omega_\mathrm{c} \hat{c}^\dagger \hat{c} 
+\hbar \sum_{i=1}^N \left( J_\mathrm{R}|\mathrm{R}_i\rangle \langle \mathrm{R}_i| + J_\mathrm{P}|\mathrm{P}_i\rangle \langle \mathrm{P}_i|\right)  \left( \hat{c}^\dagger +\hat{c} \right)^2.
\end{align}
According to the Thomas-Reiche-Kuhn sum rule~\cite{Nataf10, Liberato14, Kockum19}, $J_{\mathrm{R},\mathrm{P}}=g_{\mathrm{R},\mathrm{P}}^2/\omega_{\mathrm{R},\mathrm{P}}$. 
In the alternative dipole gauge, which can be obtained from the Coulomb gauge by making the G\"{o}ppert-Mayer gauge transformation, in addition to the conventional dipole interaction, the so-called dipole self-energy also exists, which is comparable in magnitude with the dipole interaction in the ultrastrong coupling regime and needs to be considered~\cite{Stefano19}.

Using the Fermi's golden rule, the reaction rate is obtained as 
\begin{align}
k=N\sum_\mu f_\mu \sum_\nu  k_{\mu\to \nu},
\label{eq: total rate}
\end{align}
where $\mu$ and $\nu$ label all energy eigenstates of the hybrid molecule-plus-cavity system in the initial configuration, in which all molecules are in the R state, and in the final configuration, in which one molecule changes to the P state and the others remain in the R state, respectively. The factor of $N$ in Eq.~\eqref{eq: total rate} accounts for the fact that each molecule can make a transition from R to P with equal probability. The Boltzmann distribution function for the initial state $|\mu\rangle$ is given by $f_\mu=e^{-E_\mu/k_\mathrm{B}T}/Z$, where $E_\mu$ is the energy of the $|\mu\rangle$ state, $k_\mathrm{B}$ is the Boltzmann constant, $T$ is the temperature, and $Z=\sum_\mu e^{-E_\mu/k_\mathrm{B}T}$ is the partition function. 
The rate $k_{\mu \to \nu}$ in Eq.~\eqref{eq: total rate} is given by~\cite{May-book}
\begin{align}
k_{\mu \to \nu}=&
\sqrt{\frac{\pi}{\hbar^2 k_\mathrm{B}T \lambda}} 
\left| V_{\mu, \nu} \right|^2 
\exp \left\{ - \frac{(\Delta E_{\mu,\nu}-\lambda)^2}{4\lambda k_\mathrm{B}T}\right\},
\label{eq: rate between mu and nu states}
\end{align}
where $\Delta E_{\mu,\nu}=E_\mu-E_\nu$ is the energy difference between the initial state $|\mu\rangle$ and the final state $|\nu\rangle$. 
The right-hand side of Eq.~\eqref{eq: rate between mu and nu states} has a similar form to the electron transfer rate given in Marcus-Levich-Jortner theory except that the molecular vibrational states are replaced by the energy eigenstates $\mu$ and $\nu$ of the molecule-plus-cavity hybrid system.
Under the Condon approximation, the coupling $V_{\mu,\nu}$ can be expressed in terms of the Franck-Condon factor as $V_{\mu,\nu}=V_\mathrm{RP} \langle \chi_\mu| \chi_\nu \rangle$, where $\chi_\mu$ and $\chi_\nu$ represent the nuclear wavefunctions of the $|\mu\rangle$ and $|\nu\rangle$ states along the high-frequency vibrational mode.

\section{Single-molecule system}
\label{sec: Single-molecule system}
\begin{figure}[tbp] 
  \centering
  \includegraphics[keepaspectratio]{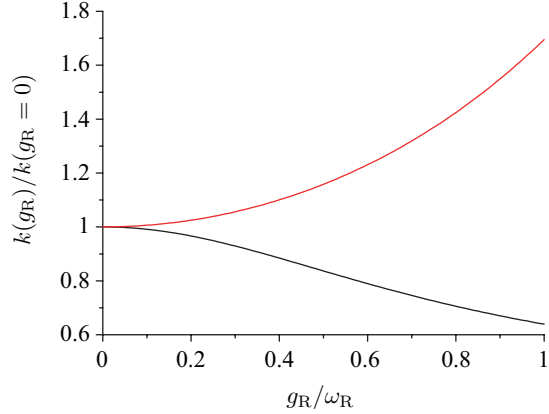}
  \caption{Relative change of the reaction rate $k(g_\mathrm{R})/k(g_\mathrm{R}=0)$ as a function of the coupling strength $g_\mathrm{R}$ (normalized by the vibrational frequency $\omega_\mathrm{R}$) for the two different cases of $g_\mathrm{P}=g_\mathrm{R}$ (black) and $g_\mathrm{P}=0$ (red).}
  \label{fig: the relative change of the reaction rate as a function of the coupling strength}
\end{figure}

A single-molecule system, i.e., with $N=1$, is considered. 
To investigate the modification of the reaction rate by coupling the molecular vibration to the cavity mode, the reaction rate given by Eq.~\eqref{eq: total rate} is calculated for a variable coupling strength $g_\mathrm{R}$. 
Two different situations, in which the coupling strengths $g_\mathrm{R}$ and $g_\mathrm{P}$ are nearly equal to each other in one case and differ significantly in the other case, are examined. 
Figure~\ref{fig: the relative change of the reaction rate as a function of the coupling strength} shows the relative change $k(g_\mathrm{R})/k(g_\mathrm{R}=0)$ of the reaction rate as a function of the normalized coupling strength $0\leq g_\mathrm{R}/\omega_\mathrm{R} \leq 1$ for the two different cases of $g_\mathrm{P}=g_\mathrm{R}$ and $g_\mathrm{P}=0$. 
Here, both of the energy difference $\Delta E_\mathrm{RP}$ and the detuning $\delta=\omega_\mathrm{c}-\omega_\mathrm{R}$ are set to be zero. 
The other parameters of the system are $\omega_\mathrm{R}=\omega_\mathrm{P}=1000\,\text{cm}^{-1}$ corresponding to the typical order of magnitude of molecular vibration frequency, $\lambda=0.5\,\text{eV}$, and $T=300\,\text{K}$. 
The reorganization energy associated with the high-frequency vibrational mode is set to be $\lambda_\mathrm{v}=\hbar\omega_\mathrm{R}d_\mathrm{RP}^2/2=0.5\,\text{eV}$, which corresponds to $d_\mathrm{RP}\simeq 2$ for $\omega_\mathrm{R}=1000\,\text{cm}^{-1}$. 
From Fig.~\ref{fig: the relative change of the reaction rate as a function of the coupling strength}, it is evident that up to the coupling strength $g_\mathrm{R}\lesssim \omega_\mathrm{R}$, which is the largest coupling strength realized in experiments thus far ($\simeq740$ cm$^{-1}$ in Ref.~\cite{Hiura19}), the reaction rate can either increase or decrease by coupling the molecular vibration to the cavity mode. 
The decrease in the reaction rate is observed when the coupling strengths $g_\mathrm{R}$ and $g_\mathrm{P}$ for the reactant and product states are close to each other; the reverse is true when the coupling strengths differ significantly from each other. 
It should, however, be expected that for sufficiently stronger couplings, the reaction rate would increase regardless of the relative coupling strength $g_\mathrm{P}/g_\mathrm{P}$, owing to the excitation of a large fraction of high-energy vibrational states.

At first sight, the decrease in the reaction rate by the molecule-cavity coupling is counterintuitive as it is expected that the excitation of high-energy vibrational states in the molecule would facilitate the chemical reaction. To obtain physical insights into this behavior of the reaction rate, we performed the following analysis. In the case of equal coupling strengths $g_\mathrm{R}=g_\mathrm{P}$ and zero energy difference, $\Delta E_\mathrm{RP}=0$, the energy shifts induced by the molecule-cavity coupling are the same for the reactant and product states. As a result, from Eq.~\eqref{eq: rate between mu and nu states}, the modification of the reaction rate can primarily be attributed to the change in the Franck-Condon factor due to the mixing of ground and excited states of molecular vibration in the ground state of the hybrid system. Indeed, the ground states $|\mu_\mathrm{R}=0\rangle$ and $|\nu_\mathrm{P}=0\rangle$ formed by the coupling of the molecular vibration to the cavity mode when the molecule is in the reactant and the product states, respectively, can be expanded in terms of the molecule's vibrational states and the photonic Fock states as
\begin{align}
|\mu_\mathrm{R}=0\rangle=&\,c_0|n_\mathrm{R}=0, n_\mathrm{c}=0\rangle + c_1|n_\mathrm{R}=1, n_\mathrm{c}=1\rangle \nonumber\\
&+c_2|n_\mathrm{R}=2, n_\mathrm{c}=0\rangle + c_3 |n_\mathrm{R}=0, n_\mathrm{c}=2\rangle \nonumber\\
&+\cdots
\label{eq: expansion of the hybrid system's ground state 1} 
\end{align}
and
\begin{align}
|\nu_\mathrm{P}=0\rangle=&\,c_0|n_\mathrm{P}=0, n_\mathrm{c}=0\rangle + c_1|n_\mathrm{P}=1, n_\mathrm{c}=1\rangle \nonumber\\
&+c_2|n_\mathrm{P}=2, n_\mathrm{c}=0\rangle + c_3 |n_\mathrm{P}=0, n_\mathrm{c}=2\rangle \nonumber\\
&+\cdots,
\label{eq: expansion of the hybrid system's ground state 2}
\end{align}  
where $n_\mathrm{R,P}$ and $n_\mathrm{c}$ denote the number of molecular vibration's quanta and that of cavity photons, respectively. It can be seen from Eqs.~\eqref{eq: expansion of the hybrid system's ground state 1} and \eqref{eq: expansion of the hybrid system's ground state 2} that only states with an even total number of vibrational and photonic quanta appear in the expansions of the ground states $|\mu_\mathrm{R}=0\rangle$ and $|\nu_\mathrm{P}=0\rangle$ of the hybrid system. This is because of the structure of the interaction Hamiltonian~\eqref{eq: interaction Hamiltonian} in which particles are either created or annihilated in pairs. 

More importantly, the coefficients $c_0$ and $c_2$ in the expansions~\eqref{eq: expansion of the hybrid system's ground state 1} and \eqref{eq: expansion of the hybrid system's ground state 2} have opposite signs. This can be understood by the perturbation theory, which is valid in the perturbation regime, as stemming from the negative factor of $(i\hbar g_\mathrm{R,P})^2<0$, which is the product of the amplitudes of two operators $\hat{a}_\mathrm{R,P}^\dagger \hat{c}^\dagger$ and $\hat{a}_\mathrm{R,P}^\dagger \hat{c}$ that connect the unperturbed ground state $|n_\mathrm{R,P}=0, n_\mathrm{c}=0\rangle$ to the vibrational excited state $|n_\mathrm{R,P}=2, n_\mathrm{c}=0\rangle$ (see Supplementary Information for details). 
However, it turns out that this is still valid for the ultrastrong coupling regime $g_\mathrm{R,P}\sim \omega_\mathrm{R,P}\sim \omega_\mathrm{c}$ under consideration. 
Numerically, we found that $c_0\simeq 0.91$ and $c_2\simeq -0.19$ for $g_\mathrm{R,P}=\omega_\mathrm{R,P}$. 
It can be seen from Eqs.~\eqref{eq: expansion of the hybrid system's ground state 1} and \eqref{eq: expansion of the hybrid system's ground state 2} that the modification of the Franck-Condon factor due to the change of the ground state of the hybrid system from $|n_\mathrm{R}=0\rangle$ ($|n_\mathrm{P}=0\rangle$) to $|\mu_\mathrm{R}=0\rangle$ ($|\nu_\mathrm{R}=0\rangle$) is mainly determined by the factor of $\langle n_\mathrm{R}=2|n_\mathrm{P}=0\rangle=\langle n_\mathrm{R}=0|n_\mathrm{P}=2\rangle$ multiplied by the coefficient $c_0c_2$. 
On the other hand, the factor of $\langle n_\mathrm{R}=2|n_\mathrm{P}=0\rangle=\langle n_\mathrm{R}=0|n_\mathrm{P}=2\rangle$ is a positive real number for a broad range of parameters around $\omega_\mathrm{R}\simeq \omega_\mathrm{P}$. This can be deduced from the following relations between the Franck-Condon factors~\cite{May-book}
\begin{align}
\langle n_\mathrm{R}=0|n_\mathrm{P}=1\rangle=&\,-\frac{d_\mathrm{RP}\sqrt{2\epsilon}}{1+\epsilon}\langle n_\mathrm{R}=0|n_\mathrm{P}=0\rangle,\\
\langle n_\mathrm{R}=0|n_\mathrm{P}=2\rangle=&\,\frac{1-\epsilon}{\sqrt{2}(1+\epsilon)}\langle n_\mathrm{R}=0|n_\mathrm{P}=0\rangle \nonumber\\
&-\frac{d_\mathrm{RP}\sqrt{\epsilon}}{1+\epsilon}\langle n_\mathrm{R}=0|n_\mathrm{P}=1\rangle,\\
\langle n_\mathrm{R}=1|n_\mathrm{P}=0\rangle=&\,\frac{2\sqrt{2}d_\mathrm{RP}}{1+\epsilon}\langle n_\mathrm{R}=0|n_\mathrm{P}=0\rangle,\\
\langle n_\mathrm{R}=2|n_\mathrm{P}=0\rangle=&\,\frac{\epsilon-1}{\sqrt{2}(1+\epsilon)}\langle n_\mathrm{R}=0|n_\mathrm{P}=0\rangle \nonumber\\
&+\frac{2d_\mathrm{RP}}{1+\epsilon}\langle n_\mathrm{R}=1|n_\mathrm{P}=0\rangle,
\end{align}
where $\epsilon \equiv \omega_\mathrm{R}/\omega_\mathrm{P}$. 
Combining the facts that $c_0c_2<0$ and $\langle n_\mathrm{R}=2|n_\mathrm{P}=0\rangle=\langle n_\mathrm{R}=0|n_\mathrm{P}=2\rangle>0$, we obtain a negative change in the value of the Franck-Condon factor due to the mixing of molecular vibration's ground and excited states in the ground state of the hybrid system. 
This, in turn, results in a decrease in the reaction rate by coupling the molecular vibration to the cavity mode.

\begin{figure}[tbp] 
  \centering
  \includegraphics[keepaspectratio]{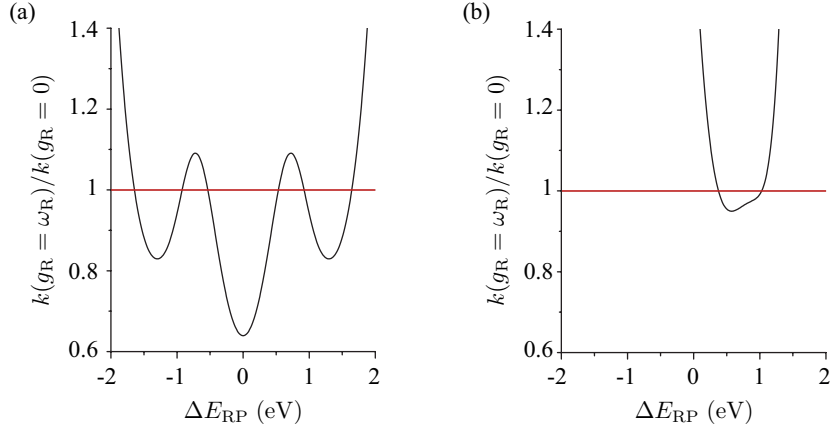}
  \caption{Relative change of the reaction rate $k(g_\mathrm{R}=\omega_\mathrm{R})/k(g_\mathrm{R}=0)$ by the molecule-cavity coupling as a function of the energy difference $\Delta E_\mathrm{RP}$ between the reactant and product states. (a) $g_\mathrm{P}=g_\mathrm{R}$. (b) $g_\mathrm{P}=0$. The red line shows the value of the reaction rate for the bare molecule as a guide for the eyes.}
  \label{fig: dependence of the reaction rate on the energy difference between the reactant and product states}
\end{figure}

To investigate the effect of energy resonance, we calculated the relative change of the reaction rate $k(g_\mathrm{R}=\omega_\mathrm{R})/k(g_\mathrm{R}=0)$ for a variable energy difference $\Delta E_\mathrm{RP}$. 
The result is shown in Fig.~\ref{fig: dependence of the reaction rate on the energy difference between the reactant and product states}a for the case that coupling strengths $g_\mathrm{R}$ and $g_\mathrm{P}$ are equal and in Fig.~\ref{fig: dependence of the reaction rate on the energy difference between the reactant and product states}b for the case of $g_\mathrm{P}=0$. 
In either case, both regions of $\Delta E_\mathrm{RP}$ exist for which the reaction rate increases or decreases by the molecule-cavity coupling. 
However, the region of $\Delta E_\mathrm{RP}$ for which the reaction rate decreases, i.e., $k(g_\mathrm{R}=\omega_\mathrm{R})<k(g_\mathrm{R}=0)$, is much broader in the case of $g_\mathrm{R}=g_\mathrm{P}$ than in the case of $g_\mathrm{P}=0$. 
The amount of decrease of the reaction rate is also much larger in the case of $g_\mathrm{R}=g_\mathrm{P}$. 
In particular, in the case of $g_\mathrm{R}=g_\mathrm{P}$ the minimum of the reaction rate is located at $\Delta E_\mathrm{RP}=0$, regardless of the detailed values of the other parameters of the system, as opposed to the case of $g_\mathrm{P}=0$. 
This reflects that the modification of the reaction rate by the molecule-cavity coupling is mainly governed by the mixing of ground and excited states of molecular vibration in the ground state of the hybrid system if the coupling strengths $g_\mathrm{R}$ and $g_\mathrm{P}$ are close to each other. 
In contrast, if the two coupling strengths differ significantly from each other, the modification of the reaction rate is mainly attributed to the relative shifts of the energy levels in the system induced by the molecule-cavity coupling. 
This is also supported by the fact that the value of $\Delta E_\mathrm{RP}$ at the minimum of the reaction rate in Fig.~\ref{fig: dependence of the reaction rate on the energy difference between the reactant and product states}b is close to the resonance peak of the reaction rate $k(g_\mathrm{R}=0)$ for the bare molecule (see Supplementary Information).

\begin{figure}[tbp] 
  \centering
  \includegraphics[keepaspectratio]{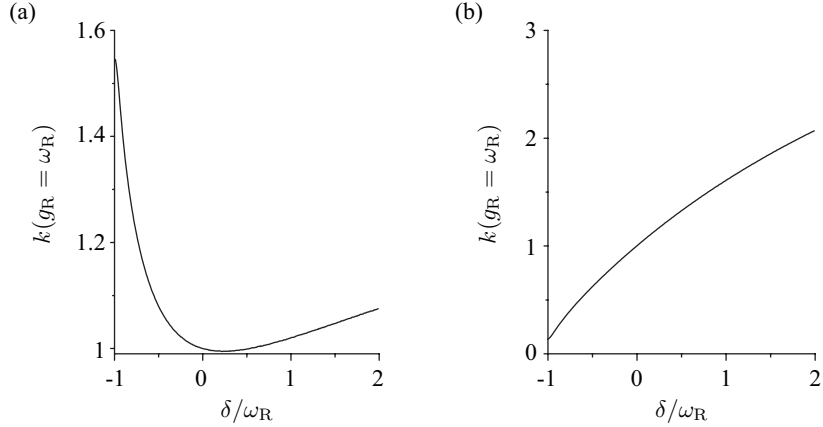}
  \caption{Dependence of the reaction rate $k(g_\mathrm{R}=\omega_\mathrm{R})$ of the coupled molecule-cavity system on the detuning $\delta=\omega_\mathrm{c}-\omega_\mathrm{R}$ of the cavity frequency relative to the molecule's vibrational frequency (normalized by $\omega_\mathrm{R}$) for the case of $g_\mathrm{P}=g_\mathrm{R}$ (a) and $g_\mathrm{P}=0$ (b). Here the reaction rate is normalized by its value at zero detuning $\delta=0$.}
  \label{fig: dependence of the reaction rate on the detuning}
\end{figure}

We also investigated the dependence of the reaction rate $k(g_\mathrm{R}=\omega_\mathrm{R})$ of the coupled molecule-cavity system on the detuning $\delta=\omega_\mathrm{c}-\omega_\mathrm{R}$ of the cavity frequency relative to the molecule's vibrational frequency (while keeping the energy difference $\Delta E_\mathrm{RP}=0$ constant). 
The result is shown in Fig.~\ref{fig: dependence of the reaction rate on the detuning}a and Fig.~\ref{fig: dependence of the reaction rate on the detuning}b for the cases of $g_\mathrm{P}=g_\mathrm{R}$ and $g_\mathrm{P}=0$, respectively. 
It can be seen that there is a broad dip near $\delta=0$ in the case of $g_\mathrm{P}=g_\mathrm{R}$. 
In this case the reaction rate decreases by the molecule-cavity coupling (Fig.~\ref{fig: the relative change of the reaction rate as a function of the coupling strength}). 
In contrast, in the case of $g_\mathrm{P}=0$, for which the reaction rate increases by the molecule-cavity coupling (Fig.~\ref{fig: the relative change of the reaction rate as a function of the coupling strength}), no peak is observed around $\delta=0$. 
The absence of a sharp resonance behavior here can be attributed to the system being in the ultrastrong coupling regime, where the coupling strength is comparable in magnitude with the other characteristic energies of the system and thus can compensate for a large energy detuning. For a comparison with experimental results, it should be noted that except for Ref.~\cite{Hiura19} where $g\simeq 740$ cm$^{-1}$, all other experiments of molecular vibration polariton were not in the ultrastrong coupling regime as the Rabi frequency is less than 100 cm$^{-1}$, an order of magnitude smaller than the vibrational frequency. Due to such a small ratio of the coupling strength to the vibrational frequency, a sharp resonance was observed in these experiments~\cite{Thomas16, Thomas19, Lather19}. It should be expected that by going closer to the ultrastrong coupling regime $g\simeq \omega$, the broadening of the resonance would be observed in experiments.  

\section{$N$-molecule system}
\label{sec: Multi-molecule system}
To investigate the collective effect in a system of identical molecules coupled to a common mode of an optical cavity, we numerically calculated the reaction rate for a system of $N=2$ molecules.
It should be noted that in this paper we consider only the case that the reaction occurs within each molecule, namely an intramolecular process rather than an intermolecular one. Despite that, since the formation of molecular polariton involves a superposition of vibrational excitations of all molecules coupled to the cavity (see discussion below), the effect of molecular polariton formation on the reaction rate of a many-molecule system is not simply a sum of its effect on each molecule. 
The obtained dependences of the reaction rate on the coupling strength $g_\mathrm{R}$, the energy difference $\Delta E_\mathrm{RP}$ and the detuning $\delta=\omega_\mathrm{c}-\omega_\mathrm{R}$ for $N=2$ are, however, qualitatively similar to those of the single-molecule system (see Supplementary Information).  

\begin{figure}[tbp] 
  \centering
  \includegraphics[keepaspectratio]{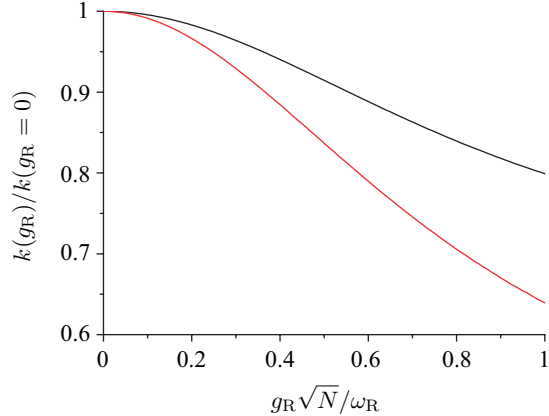}
  \caption{Relative change of the reaction rate $k(g_\mathrm{R})/k(g_\mathrm{R}=0)$ as a function of the collective Rabi frequency (or coupling strength) $g_\mathrm{R}\sqrt{N}$ (normalized by the vibrational frequency $\omega_\mathrm{R}$) for the case of $g_\mathrm{P}=g_\mathrm{R}$ in a system of $N=2$ identical molecules (black) and in the single-molecule system (red).}
  \label{fig: compare the reaction rate between single and multi-molecule systems}
\end{figure}

To examine the quantitative difference between the multi-molecule system and the single-molecule system, we compared the relative change $k(g_\mathrm{R})/k(g_\mathrm{R}=0)$ of the reaction rate as a function of the normalized coupling strength $0\leq g_\mathrm{R}\sqrt{N}/\omega_\mathrm{R} \leq 1$ for the case of $g_\mathrm{P}=g_\mathrm{R}$ and $\Delta E_\mathrm{RP}=0$ in the two systems. 
Here, $g_\mathrm{R}\sqrt{N}$ is the collective Rabi frequency (or coupling strength) and it reduces to the single-emitter Rabi frequency (or coupling strength) $g_\mathrm{R}$ if $N=1$. 
The result is shown in Fig.~\ref{fig: compare the reaction rate between single and multi-molecule systems}. 
It is evident that in both systems, the reaction rate decreases by coupling the molecular vibration to the optical cavity mode. However, the amount of decrease is smaller in the system of $N=2$ molecules than in the single-molecule system by a factor of approximately 1/2. 

To obtain physical insights into this behavior, we examined how the Franck-Condon factor between the ground states of the hybrid molecule-plus-cavity system in the initial and final configurations changes with the variation of the number of molecules. 
As shown in the single-molecule system, when the coupling strengths $g_\mathrm{R}$ and $g_\mathrm{P}$ are close to each other and $\Delta E_\mathrm{RP}\simeq 0$, the modification of the reaction rate by the molecule-cavity coupling is mainly determined by the change of the Franck-Condon factor between the ground states of the hybrid system in the initial and final configurations. 
For the system of identical molecules, the initial configuration consists of all molecules in the reactant states while in the final configuration one molecule changes to the product state. 

If a system of $N$ identical molecules is coupled to a common optical cavity mode, only a single molecular collective mode fully symmetric with respect to the exchange of any pair of molecules is coupled to the optical cavity, while a total number of $N-1$ other molecular collective modes are not coupled to the cavity mode as they are in dark states. 
The fully symmetric molecular collective mode for the initial configuration, in which all molecules are in the reactant state, is given by $\hat{A}_\mathrm{i}^+=(1/\sqrt{N})\sum_{j=1}^N \hat{a}_\mathrm{R}^{(j)}$, while that for the final configuration with one molecule, say the $N$th molecule being in the product state, is given by $\hat{A}_\mathrm{f}^+=(1/\sqrt{N})\left[\sum_{j=1}^{N-1}\hat{a}_\mathrm{R}^{(j)}+\hat{a}_\mathrm{P}^{(N)}\right]$. 
The other $N-1$ collective molecular modes are denoted by a set of annihilation operators $\{\hat{A}_{\mathrm{i},\mathrm{f}}^-\}$.
The ground states $|\mu_\mathrm{i}=0\rangle$ and $|\nu_\mathrm{f}=0\rangle$ of the hybrid molecule-plus-cavity system in the initial and final configurations, respectively, can be expanded in terms of the molecular and photonic Fock states as
\begin{align}
|\mu_\mathrm{i}=0\rangle=&c_0|n_{A_\mathrm{i}^+}=0, n_\mathrm{c}=0, \{n_{A_\mathrm{i}^-}=0\}\rangle \nonumber\\
&+c_1|n_{A_\mathrm{i}^+}=1, n_\mathrm{c}=1, \{n_{A_\mathrm{i}^-}=0\}\rangle \nonumber\\
&+c_2|n_{A_\mathrm{i}^+}=2, n_\mathrm{c}=0, \{n_{A_\mathrm{i}^-}=0\}\rangle \nonumber\\
&+c_3|n_{A_\mathrm{i}^+}=0, n_\mathrm{c}=2, \{n_{A_\mathrm{i}^-}=0\}\rangle \nonumber\\
&+\cdots
\end{align}
and
\begin{align}
|\nu_\mathrm{f}=0\rangle=&c_0|n_{A_\mathrm{f}^+}=0, n_\mathrm{c}=0, \{n_{A_\mathrm{f}^-}=0\}\rangle \nonumber\\
&+c_1|n_{A_\mathrm{f}^+}=1, n_\mathrm{c}=1, \{n_{A_\mathrm{f}^-}=0\}\rangle \nonumber\\
&+c_2|n_{A_\mathrm{f}^+}=2, n_\mathrm{c}=0, \{n_{A_\mathrm{f}^-}=0\}\rangle \nonumber\\
&+c_3|n_{A_\mathrm{f}^+}=0, n_\mathrm{c}=2, \{n_{A_\mathrm{f}^-}=0\}\rangle \nonumber\\
&+\cdots,
\end{align}
where the coefficients $c_{0,1,2,3}$ are the same as those in Eqs.~\eqref{eq: expansion of the hybrid system's ground state 1} and \eqref{eq: expansion of the hybrid system's ground state 2} as long as the collective coupling strength $g_\mathrm{R}\sqrt{N}$ is kept fixed for variable $N$. 
Similar to the single-molecule system, the modification of the Franck-Condon factor due to the change of the ground state of the hybrid system is mainly determined by the factors of $\langle n_{A_\mathrm{i}^+}=0, n_\mathrm{c}=0, \{n_{A_\mathrm{i}^-}=0\}|n_{A_\mathrm{f}^+}=2, n_\mathrm{c}=0, \{n_{A_\mathrm{f}^-}=0\}\rangle$ and $\langle n_{A_\mathrm{i}^+}=2, n_\mathrm{c}=0, \{n_{A_\mathrm{i}^-}=0\}| n_{A_\mathrm{f}^+}=0, n_\mathrm{c}=0, \{n_{A_\mathrm{f}^-}=0\}\rangle$ multiplied by the coefficient $c_0c_2$.
Using the expansion of the fully symmetric molecular collective modes $\hat{A}_{\mathrm{i},\mathrm{f}}^+$ in terms of the single-molecule operators $\hat{a}_{\mathrm{R},\mathrm{P}}^{(j)}$ ($j=1,\cdots,N$), it is found that
\begin{widetext}
\begin{align}
\langle n_{A_\mathrm{i}^+}=0, n_\mathrm{c}=0, \{n_{A_\mathrm{i}^-}=0\}|n_{A_\mathrm{f}^+}=0, n_\mathrm{c}=0, \{n_{A_\mathrm{f}^-}=0\}\rangle=&\langle n_\mathrm{R}=0|n_\mathrm{P}=0 \rangle, \\
\langle n_{A_\mathrm{i}^+}=0, n_\mathrm{c}=0, \{n_{A_\mathrm{i}^-}=0\}|n_{A_\mathrm{f}^+}=2, n_\mathrm{c}=0, \{n_{A_\mathrm{f}^-}=0\}\rangle=&\frac{1}{N}\langle n_\mathrm{R}=0|n_\mathrm{P}=2 \rangle, \\
\langle n_{A_\mathrm{i}^+}=2, n_\mathrm{c}=0, \{n_{A_\mathrm{i}^-}=0\}| n_{A_\mathrm{f}^+}=0, n_\mathrm{c}=0, \{n_{A_\mathrm{f}^-}=0\}\rangle=&\frac{1}{N}\langle n_\mathrm{R}=2|n_\mathrm{P}=0 \rangle.
\end{align}
\end{widetext}
It is clear that compared to the single-molecule system, there appears an additional factor of $1/N$ in the expressions of $\langle n_{A_\mathrm{i}^+}=0, n_\mathrm{c}=0, \{n_{A_\mathrm{i}^-}=0\}|n_{A_\mathrm{f}^+}=2, n_\mathrm{c}=0, \{n_{A_\mathrm{f}^-}=0\}\rangle$ and $\langle n_{A_\mathrm{i}^+}=2, n_\mathrm{c}=0, \{n_{A_\mathrm{i}^-}=0\}| n_{A_\mathrm{f}^+}=0, n_\mathrm{c}=0, \{n_{A_\mathrm{f}^-}=0\}\rangle$. 
As a result, the change of the Franck-Condon factor between the ground states of the hybrid system in the initial and final configurations is given by 
\begin{align}
\langle \mu_\mathrm{i}=0|\nu_\mathrm{f}=0\rangle\simeq &\,c_0^2 \langle n_\mathrm{R}=0|n_\mathrm{P}=0\rangle \nonumber\\
&+\frac{2c_0c_2}{N}\langle n_\mathrm{R}=0|n_\mathrm{P}=2 \rangle,
\end{align}
where we used the fact that $\langle n_\mathrm{R}=0|n_\mathrm{P}=2 \rangle=\langle n_\mathrm{R}=2|n_\mathrm{P}=0 \rangle$ for $\omega_\mathrm{R}=\omega_\mathrm{P}$. 
Since $|c_2|\ll |c_0|$ for $g_\mathrm{R}\sqrt{N}\lesssim \omega_\mathrm{R}$ (see Sec.~\ref{sec: Single-molecule system}), the change of the square of the Franck-Condon factor appearing in Eq.~\eqref{eq: rate between mu and nu states} for the reaction rate is found to be
\begin{align}
\langle \mu_\mathrm{i}=0|\nu_\mathrm{f}=0\rangle^2\simeq&\,
c_0^4 \langle n_\mathrm{R}=0|n_\mathrm{P}=0\rangle^2 +\frac{2c_0^3c_2}{N} \nonumber\\
&\times\langle n_\mathrm{R}=0|n_\mathrm{P}=0\rangle 
\langle n_\mathrm{R}=0|n_\mathrm{P}=2 \rangle.
\end{align}
It is evident that compared to the single-molecule system the decrease ($c_0c_2<0$) of the reaction rate by the molecule-cavity coupling is reduced approximately by a factor of $1/N$. This is in agreement with the numerical result shown in Fig.~\ref{fig: compare the reaction rate between single and multi-molecule systems} for $N=2$. 
Consequently, it should be expected that the effect on the reaction rate of the mixing of ground and excited states of molecular vibration in the ground state of the hybrid system, through which the Franck-Condon factor between the initial and final states of the transition is altered, would be large for a system with a few molecules but would decrease to extremely small in a system containing a large number of molecules.

On the other hand, if the coupling strengths $g_\mathrm{R}$ and $g_\mathrm{P}$ for the reactant and product states differ significantly from each other, the relative shift of the energy levels induced by the molecule-cavity coupling should play the dominant role in the modification of the reaction rate, as already demonstrated for the single-molecule system. In the case of a collection of $N$ identical molecules coupled to a common cavity mode, the energy shift of the formed polariton state relative to the bare molecular energy is given by the collective Rabi frequency (or coupling strength) $g_{\mathrm{R},\mathrm{P}}\sqrt{N}$. Therefore, as long as the collective coupling strength is kept fixed as the number of molecules is varied (i.e., the single-emitter coupling strength $g_\mathrm{R,P}$ is proportional to $1/\sqrt{N}$), it can be expected that the effect of the energy shifts in the system on the reaction rate would not disappear in a system containing a large number of molecules~\cite{Angulo19}.

\section{Discussion}
\label{sec: Discussion}
We investigated how the rate of an electron-transfer reaction is modified by coupling the molecular vibrations to an optical cavity mode in the ultrastrong coupling regime, where the coupling strength is comparable in magnitude with both the vibrational and cavity frequencies.
It was found that the modification of the reaction rate is determined by two main factors: the relative shifts of the energy levels induced by the molecule-cavity coupling and the mixing of ground and excited states of molecular vibration in the ground state of the hybrid system, through which the Franck-Condon factor between the initial and final states of the transition is altered. 
The former factor dominates if there is a significant difference between the molecule-cavity coupling strengths for the reactant and product states, giving rise to an increase in the reaction rate over a wide range of system's parameters. 
In contrast, if the coupling strengths and energy levels of the reactant and product states are close to each other, the latter factor becomes predominant and counterintuitively leads to a decrease in the reaction rate. 

It is noteworthy that the mixing of ground and excited states of molecular vibration in the ground state of the hybrid system is an effect unique to the ultrastrong coupling regime, as opposed to the effect of energy shift which exists in both strong and ultrastrong coupling regimes.
The effect of the mixing of vibrational excitations on the reaction rate is, however, suppressed in a system containing a large number of molecules due to the collective nature of the resulting polariton. Therefore, this effect should be observed in a system containing a small number of molecules. 
A strong coupling between the electronic excitation of a single molecule and an optical nanocavity has been observed in Ref.~\cite{Chikkaraddy16}. The Rabi frequency attained in this experiment ($\simeq 740$ cm$^{-1}$) is comparable in magnitude with the typical molecular vibration frequency. 
It can be expected that with the rapid advancement of experimental techniques the ultrastrong coupling for vibrational polariton will be realized in a system containing a small number of molecules, which would be a suitable platform for observing the effect of molecular vibration mixing predicted in this paper.

On the other hand, the effect of the relative shifts of the energy levels is essentially independent of the number of molecules and therefore should be dominant in a system containing a large number of molecules. Even though the result obtained in this paper cannot be used to make a direct comparison with the experimental results in Refs.~\cite{Thomas16, Thomas19, Hiura19, Lather19} because of different kinds of chemical reactions having been investigated, the result shown in Fig.~\ref{fig: dependence of the reaction rate on the energy difference between the reactant and product states} agrees at least qualitatively with what were observed in the experiments: the reaction rate can either decrease~\cite{Thomas16, Thomas19} or increase~\cite{Hiura19, Lather19}, and the amount of increase is several orders of magnitude larger than that of decrease.

The rapid progress in the realization of strong and ultrastrong couplings of both electronic and vibrational degrees of freedom of molecules to an optical cavity is expected to open new avenues regarding the physical control of various kinds of chemical reactions and dynamical processes~\cite{Ebbesen16, Ribeiro18, Feist18, Hertzog19} that, unlike other controlling approaches using intense laser fields such as the Floquet engineering~\cite{Phuc18, Phuc19b}, is based on the zero-point quantum fluctuation of the vacuum state of the cavity field.

\begin{acknowledgements}
This work was supported by JSPS KAKENHI Grant Numbers 19K14638 (N.~T.~Phuc), 17H02946 and 18H01937, and JSPS KAKENHI Grant Number 17H06437 in Innovative Areas ``Innovations for Light-Energy Conversion (I$^4$LEC)'' (A.~Ishizaki).
The computations were performed using Research Center for Computational Science, Okazaki, Japan.
\end{acknowledgements}


\end{document}